\documentclass[11pt]{article}
\usepackage[utf8]{inputenc}
\usepackage{pdflscape}
\usepackage[margin=1.0in]{geometry}
\usepackage{array}
\usepackage{setspace}
\usepackage{hyperref}
\hypersetup{colorlinks=true,linkcolor=black,citecolor=blue}
\usepackage{tikz}
\usepackage{amsmath}
\usepackage{bbold}
\usepackage{slashed}
\usepackage{enumitem}
\usepackage{physics}
\usepackage{amssymb}
\usepackage{mathtools}
\usepackage{youngtab}
\usepackage{xcolor}
\usepackage{multirow}
\usepackage{stfloats}
\usepackage{comment}

\usepackage{amsfonts}
\usepackage{graphicx}
\usepackage{subfiles}
\usepackage{mathrsfs}
\usepackage[toc,page]{appendix}

\makeatletter
\newcommand{\thickhline}{%
    \noalign {\ifnum 0=`}\fi \hrule height 1pt
    \futurelet \reserved@a \@xhline
}
\newcolumntype{"}{@{\hskip\tabcolsep\vrule width 1pt\hskip\tabcolsep}}
\makeatother
\usepackage{imakeidx}
\makeindex[columns=3, title=Alphabetical Index, intoc]
\begin{document}
\pagenumbering{gobble}

\Large
 \begin{center}
  \textbf{Spin 2 fluctuations in 1/4 BPS AdS$_3$/CFT$_2$} \\

\vskip 1cm  
\large   
{Stefano Speziali\footnote{\href{mailto:stefano.speziali6@gmail.com}{stefano.speziali6@gmail.com}} } \\

\hspace{20pt}

\small  
\it Department of Physics,\\
Swansea University,\\
Singleton Park, Swansea,\\
SA2 8PP, U.K.
\end{center}

\hspace{20pt}

\normalsize

\begin{abstract}
     In this paper we study spin 2 fluctuations around a warped $AdS_3 \times S^2 \times T^4 \times \mathcal{I}_{\rho}$ background in type IIA supergravity with small $\mathcal{N} = (0,4)$ supersymmetry. We find a class of fluctuations, which will be called \textit{universal}, that is independent of the background data and corresponds to operators with scaling dimension $\Delta = 2l +2$, being $l$ the angular-momentum-quantum-number on the $S^2$ which realises the $SU(2)_R$ symmetry. We compute the central charge for $\mathcal{N} = (0,4)$ two-dimensional superconformal theories from the action of the spin 2 fluctuations.
\end{abstract}
\newpage
\pagenumbering{arabic}
\tableofcontents

\section{Introduction}

\paragraph{} Superconformal Field Theories (SCFTs) in diverse dimensions, and with different number of supersymmetries, have been object of intense study in the past years and still constitute a rich and fruitful subject. Aside from being interesting in their own right, they play a crucial role in the AdS/CFT duality. In 1997, Maldacena \cite{Maldacena:1997re} conjectured that $d$-dimensional SCFTs are dual to $AdS_{d+1}$ backgrounds and since then the AdS/CFT duality has provided a powerful tool to make strongly coupled CFTs more tractable.

Over the years, very useful has proved to be the correspondence between SCFTs in $d>2$ with 8 Poincaré supercharges and their holographic duals. For instance, the $\mathcal{N} = 4$ three-dimensional field theories studied in \cite{Gaiotto:2008sa, Gaiotto:2008ak} have been explored from a holographic perspective in e.g. \cite{DHoker:2008lup, DHoker:2007zhm, Assel:2011xz, Lozano:2016wrs}. In four dimensions, the A-type of quivers of \cite{Gaiotto:2009we} corresponding to $\mathcal{N} = 2$ supersymmetry, already solved in \cite{Witten:1997sc}, found a holographic realisation in \cite{Gaiotto:2009gz}. Further holographic studies were given in e.g. \cite{ReidEdwards:2010qs, Aharony:2012tz, Nunez:2019gbg}. Also five dimensional SCFTs with 8 supercharges have found a holographic realisation, see for instance \cite{DHoker:2016ujz, DHoker:2016ysh, DHoker:2017mds, Bergman:2018hin} while in six dimensions $\mathcal{N} = (0,1)$ SCFTs were addressed from a QFT and holographic point of view in many papers, see for instance \cite{Apruzzi:2015wna, Gaiotto:2014lca, Cremonesi:2015bld}.

The case of two-dimensional SCFTs is particularly interesting, as they are intrinsically different from SCFTs in $d>2$. First of all, their (superconformal) algebra is infinite-dimensional \cite{Belavin:1984vu}. This makes them much more easy to analyse and, sometimes, they can even be solved exactly \cite{DiFrancesco:1997nk}. Secondly, they find many applications in string and quantum field theory, e.g. two-dimensional SCFTs make their appearance when quantising the super-Polyakov action, but they also offer a description of several critical phenomena.

In the present work we focus on two-dimensional SCFTs with $\mathcal{N} = (0,4)$ supersymmetry. Their (infinite-dimensional) superconformal algebra was constructed in \cite{Ademollo:1975an}, and studied further in the subsequent papers \cite{Ademollo:1976pp, Ademollo:1976wv}. By virtue of AdS/CFT, $\mathcal{N} = (0,4)$ two-dimensional SCFTs are supposed to be dual to type II supergravity backgrounds with an $AdS_3$ factor. In fact, an infinite family of new solutions in type IIA supergravity with an $AdS_3 \times S^2$ factor, preserving $\mathcal{N} = (0,4)$ supersymmetry, was recently built in \cite{Lozano:2019jza, Lozano:2019zvg} and further explored in \cite{Lozano:2019ywa}. All these solutions relied on the local construction given in \cite{Lozano:2019emq} which classify local solutions in massive IIA with an $AdS_3 \times S^2$ factor and an $SU(2)$ structure. The authors of \cite{Lozano:2019jza, Lozano:2019zvg, Lozano:2019ywa, Lozano:2019emq} identified the backgrounds that are dual to the IR limit of a special class of long quivers. These quivers are, in turn, made of two families of $\mathcal{N} = (4,4)$ linear quivers coupled by matter fields. We will introduce such quantum field theories and review their main features in Section \ref{section 2} and Appendix \ref{2d theories}.

An important part of the study of a class of SCFTs is the spectrum of operators, and understanding how they fit into representations of the superalgebra is a challenging and stimulating problem. In a recent work \cite{Kos:2018glc}, multiplets for the two-dimensional $\mathcal{N} = 4$ superconformal algebra have been built. These multiplets fall into short and long multiplets. The authors of \cite{Kos:2018glc} were mainly interested in applications of two-dimensional $\mathcal{N} = 4$ superconfomal algebra to a numerical bootstrap study. Here, we will rely on their results concerning representations of the $\mathcal{N} = 4$ superconfomal algebra to study holographically the spectrum of operators.

In AdS/CFT, the linearised fluctuations of the supergravity background capture the spectrum of the dual gauge-invariant superconformal operators. Therefore, the main motivation of the present work is to study (some) fluctuations around the backgrounds first presented in \cite{Lozano:2019jza}, in order to holographically reproduce (some of) the spectrum of operators already found in \cite{Kos:2018glc}. Constructing linearised fluctuations for the full supergravity background is not an easy task\footnote{See the seminal papers \cite{PhysRevD.32.389, Deger:1998nm} where the full Kaluza-Klein spectrum was obtained for the non-warped cases of $AdS_5 \times S^5$ and $AdS_3 \times S^3$.}. However, as noticed in \cite{Bachas:2011xa}, for the case of sole spin 2 fluctuations the problem simplifies considerably. It turns out that spin 2 fluctuations, which are given in terms of perturbations of the backgroud metric, solve an equation that depends only on the underlying geometry of the background. This strategy has been applied succesfully in e.g. \cite{Klebanov:2009kp, Schmude:2016bqp, Pang:2017omp, Richard:2014qsa, Chen:2019ydk, Itsios:2019yzp, Gutperle:2018wuk, Passias:2018swc, Passias:2016fkm} for the case of four-, five- and six-dimensional SCFTs. We will follow a similar path for the case of a warped $AdS_3$.

The paper is organised as follows. In Section \ref{section 2} we briefly review the background presented in \cite{Lozano:2019jza} along with the dual CFT interpretation further explored in \cite{Lozano:2019zvg, Lozano:2019ywa}. In Section \ref{section fluctuations} we derive an equation for spin 2 fluctuations of the metric background. These are transverse and traceless fluctuations along the $AdS_3$ part of the geometry and correspond to massive rank-2 tensors. In Section \ref{section universal solution} we identify a particular class of solutions. These are the \textit{universal} type of solutions, as they are independent of the background data. As we will see, they are also \textit{minimal} solutions as they correspond to spin 2 fluctuations for which the mass of the graviton is the minimum possible in terms of the angular momentum on the $S^2$.  In Section \ref{section implications for field theory} we discuss the implications for the dual field theory. In particular, we will see that the universal solution corresponding to massless gravitons is dual to the energy momentum tensor operator of the dual field theory. Finally, in Section \ref{section central charge}, we will see how to compute the central charge for the $\mathcal{N} = (0,4)$ long quiver of \cite{Lozano:2019zvg} from the action of the spin 2 fluctuations. We give our conclusions in Section \ref{section conclusions}. In the Appendices, we give an example of non universal solution (dependent on the background data) and spell out the algebra and superfield construction of $\mathcal{N} = (0,4)$ two-dimensional superconformal field theories.

\section{The gravity backgrounds and dual field theories}\label{section 2}

\paragraph{} In this section we review the global class of solutions first presented in \cite{Lozano:2019jza} as well as the proposed dual field theories. These new backgrounds are solutions to massive IIA supergravity and have the structure of a warped $AdS_3 \times S^2 \times \text{CY}_2 \times \mathcal{I}_{\rho}$, with $\mathcal{I}_{\rho}$ an interval parametrised by a coordinate labelled $\rho$.

\subsection*{The holographic backgrounds}

\paragraph{} The NS sector of the global class of solutions of \cite{Lozano:2019jza} reads
\begin{equation}\label{eq:class I background}
\begin{split}
ds^2 &= \frac{u}{\sqrt{\hat{h}_4 h_8}} \left( ds^2(AdS_3) + \frac{\hat{h}_4 h_8}{4 \hat{h}_4 h_8 + (u')^2} ds^2(S^2) \right) + \sqrt{\frac{\hat{h}_4}{h_8}} ds^2(\text{CY}_2) + \frac{\sqrt{\hat{h}_4 h_8}}{u} d \rho^2 \, , \\
e^{- \Phi}&= \frac{h_8^{3/4}}{2 \hat{h}_4^{1/4} \sqrt{u}} \sqrt{4 \hat{h}_4 h_8 + (u')^2} \, , \quad H = \frac{1}{2}d \left( - \rho + \frac{u u'}{4 \hat{h}_4 h_8 + (u')^2} \right) \wedge \text{Vol}(S^2) + \frac{1}{h_8} d \rho \wedge H_2 \, .
\end{split}
\end{equation}
Here $\Phi$ is the dilaton, $H$ the NS three-form and the metric is given in string frame. $H_2$ is a two form whose explicit form was given in \cite{Lozano:2019emq}. The functions $u$, $\hat{h}_4$, $h_8$ are functions only of the $\rho$ coordinate.\footnote{A complication of this system is when $\hat{h}_4$ has support on $(\rho, \text{CY}_2)$. The more general backgrounds deriving from this assumption are discussed in the original paper \cite{Lozano:2019emq}.} A prime denotes a derivative with respect to $\rho$.

The RR sector reads
\begin{equation}\label{eq:class I background RR}
\begin{split}
F_0 &= h_8' \, , \quad F_2 = - H_2 - \frac{1}{2} \left( h_8 - \frac{h_8' u' u}{4 h_8 \hat{h}_4+(u')^2} \right) \text{Vol}(S^2) \, , \\
F_4 &= \left( d \left( \frac{u'u}{2 \hat{h}_4} \right) + 2 h_8 d \rho \right) \wedge \text{Vol}(AdS_3) - \partial_{\rho} \hat{h}_4 \text{Vol}(\text{CY}_2) - \frac{u' u}{2(4 \hat{h}_4 h_8 + (u')^2)} H_2 \wedge \text{Vol}(S^2) \, .
\end{split}
\end{equation}
Higher RR fluxes are related to $F_0$, $F_2$ and $F_4$ as usual as $F_6 = - \star F_4$, $F_8 = \star F_2$, $F_{10} = - \star F_0$, where $\star$ is the ten-dimensional Hodge-dual operator.

It was shown in \cite{Lozano:2019emq} that supersymmetry is mantained when
\begin{equation}
u'' = 0 \, , \quad \quad H_2 + \star_{4}H_2 = 0 \, ,
\end{equation}
where $\star_{4}$ is the Hodge-dual on the CY$_2$. In the following we will consider only that class of geometries with $H_2 = 0$. Away from brane sources, the Bianchi identities imply
\begin{equation}\label{eq:Bianchi without sources}
h_8'' = 0 \, , \quad \quad \hat{h}_4'' = 0 \, . 
\end{equation}
Thus the three functions $u$, $\hat{h}_4$ and $h_8$ that appear as warping factors are at most linear in $\rho$\footnote{Again, this is true away from brane sources. In the presence of branes, the rhs' of the two equations in \eqref{eq:Bianchi without sources} receive infinite contributions in the form of delta functions. This causes $\hat{h}_4$ and $h_8$ to be piecewise linear functions.}. This will lead to considerable simplifications in the following sections.

Following \cite{Lozano:2019zvg}, we will be interested in the case of a finite interval $\mathcal{I}_{\rho}$ where both $\hat{h}_4$ and $h_8$ vanish at both ends of the interval. So, to start fixing conventions, let us set $\mathcal{I}_{\rho} = [0, \rho^*]$ and $\hat{h}_4(\bar{\rho}) = h_8(\bar{\rho}) = 0$, when $\bar{\rho}$ is equal to $0$ and  $\rho^*$. It is convenient \cite{Lozano:2019zvg} to set $\rho^* = 2 \pi (P + 1)$, with $P$ a large integer. On the other hand, $u$ vanishes only at $\rho = 0$. The general form for $\hat{h}_4$, $h_8$ and $u$ is then found to be
  \begin{equation}\label{eq:h4}
    \hat{h}_4(\rho) =
  \Upsilon  \begin{cases*}
      \frac{\beta_0}{2 \pi} \rho & $ 0 \leq \rho \leq 2 \pi$  \\
      \beta_0 + \dots + \beta_{k-1}  + \frac{\beta_k}{2 \pi}(\rho - 2 \pi k)   & $2 \pi k < \rho \leq 2 \pi (k+1) \, , \quad k = 1, \cdots, P-1$ \\
      \alpha_P - \frac{\alpha_P}{2 \pi} (\rho - 2 \pi P) & $2 \pi P < \rho \leq 2 \pi (P+1) \, , $
    \end{cases*}
\end{equation}
 \begin{equation}\label{eq:h8}
    h_8(\rho) =
    \begin{cases*}
      \frac{\nu_0}{2 \pi} \rho & $ 0 \leq \rho \leq 2 \pi$  \\
      \nu_0  + \dots + \nu_{k-1} + \frac{\nu_k}{2 \pi}(\rho - 2 \pi k)   & $2 \pi k < \rho \leq 2 \pi (k+1) \, , \quad k = 1, \cdots, P-1$ \\
      \mu_P - \frac{\mu_P}{2 \pi} (\rho - 2 \pi P) & $2 \pi P < \rho \leq 2 \pi (P+1) \, , $
    \end{cases*}
\end{equation}
and
\begin{equation}\label{eq:u}
u = \frac{b_0}{2 \pi} \rho \, .
\end{equation}
Here $\alpha_P = \sum \beta_k$ and $\mu_P = \sum \nu_k$ by continuity of $\hat{h}_4$ and $h_8$.

\subsection*{The dual field theories}

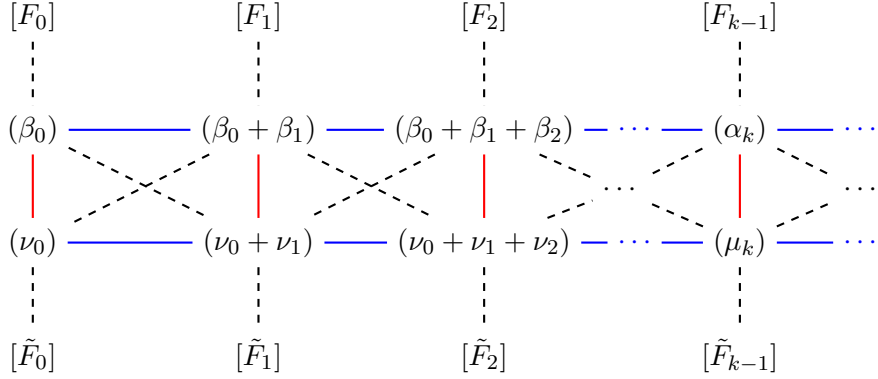
\begin{figure}[h!]
\begin{center}
\begin{tikzpicture}[
c/.style={circle, draw=black, fill=gray, very thick, minimum size=17mm},
f/.style={rectangle, draw=black, fill=white, very thick, minimum size=10mm},
]
\node at (0,0) (F0) {$[F_0]$};
\node at (3,0) (F1) {$[F_1]$};
\node at (6,0) (F2) {$[F_2]$};
\node at (9.4,0) (FK) {$[F_{k-1}]$};

\node at (0,-1.5) (b0) {$(\beta_0)$};
\node at (3,-1.5) (b1) { $(\beta_0+\beta_1)$};
\node at (6,-1.5) (b2) {$(\beta_0+\beta_1+\beta_2)$};
\node at (9.4,-1.5) (b3) {$(\alpha_k)$};

\node at (0,-3) (n0) {$(\nu_0)$};
\node at (3,-3) (n1) { $(\nu_0+\nu_1)$};
\node at (6,-3) (n2) {$(\nu_0+\nu_1+\nu_2)$};
\node at (9.4,-3) (n3) {$(\mu_k)$};

\node at (0,-4.5) (f0) {$[\tilde{F}_0]$};
\node at (3,-4.5) (f1) {$[\tilde{F}_1]$};
\node at (6,-4.5) (f2) {$[\tilde{F}_2]$};
\node at (9.4,-4.5) (fK) {$[\tilde{F}_{k-1}]$};

\draw[thick,dashed](F0)--(b0);
\draw[thick,dashed](F1)--(b1);
\draw[thick,dashed](F2)--(b2);
\draw[thick,dashed](FK)--(b3);

\draw[thick,red] (b0)--(n0);
\draw[thick,red] (b1)--(n1);
\draw[thick,red] (b2)--(n2);
\draw[thick,red] (b3)--(n3);

\node at (8,-1.5) (d) {$\bf\textcolor{blue}{\dotsb}$};

\draw[thick,blue] (b0)--(b1);
\draw[thick,blue] (b1)--(b2);
\draw[thick,blue] (b2)--(d);
\draw[thick,blue] (d)--(b3);

\node at (8,-3) (dd) {$\bf\textcolor{blue}{\dotsb}$};

\draw[thick,blue] (n0)--(n1);
\draw[thick,blue] (n1)--(n2);
\draw[thick,blue] (n2)--(dd);
\draw[thick,blue] (dd)--(n3);

\draw[thick,dashed] (n0)--(f0);
\draw[thick,dashed] (n1)--(f1);
\draw[thick,dashed] (n2)--(f2);
\draw[thick,dashed] (n3)--(fK);

\node at (7.8,-2.3) (ddd) {$\bf\textcolor{black}{\dotsb}$};

\draw[thick,dashed] (b0)--(n1);
\draw[thick,dashed] (b1)--(n0);
\draw[thick,dashed] (b1)--(n2);
\draw[thick,dashed] (b2)--(n1);
\draw[thick,dashed] (b2)--(ddd);
\draw[thick,dashed] (n2)--(ddd);
\draw[thick,dashed] (b3)--(ddd);
\draw[thick,dashed] (n3)--(ddd);

\node at (11,-1.5) (dddd) {$\bf\textcolor{blue}{\dotsb}$};
\node at (11,-3) (ddddd) {$\bf\textcolor{blue}{\dotsb}$};
\node at (11,-2.3) (dddddd) {$\bf\textcolor{black}{\dotsb}$};

\draw[thick,blue] (b3)--(dddd);
\draw[thick,blue] (n3)--(ddddd);
\draw[thick,dashed] (b3)--(dddddd);
\draw[thick,dashed] (n3)--(dddddd);

\end{tikzpicture}
\end{center}
\caption{The generic quiver whose IR is captured by the background in \eqref{eq:class I background} and \eqref{eq:class I background RR}. Each gauge node is associated with a $(4,4)$ vector multiplets. Blue lines represent $(4, 4)$ hypermultiplets. Red lines represent instead $(0, 4)$ hypermultiplets and dashed lines $(0, 2)$ Fermi multiplets.}
\label{quiver}
\end{figure}

\paragraph{} The background given in \eqref{eq:class I background}, \eqref{eq:class I background RR} with $\hat{h}_4$, $h_8$ and $u$ as in \eqref{eq:h4}, \eqref{eq:h8} and \eqref{eq:u} was found \cite{Lozano:2019zvg} to be dual to the IR limit of the quiver in Figure \ref{quiver}. More precisely, the quiver in Figure \ref{quiver} is supposed to flow in the IR to a fixed point whose dynamics is captured by the background above with the warping functions just described.

Let us spell out what the building blocks of such a quiver are.\footnote{Basics of $\mathcal{N} = (0,2)$ and $\mathcal{N} = (0,4)$ superconformal field theories in $2$ dimensions are reviewed in Appendix B. For a complete treatment see \cite{Witten:1993yc}. Very useful are also \cite{Witten:1994tz, Tong:2014yna, Putrov:2015jpa, Franco:2015tna}.}
An $SU(N)$ gauge node of the quiver is denoted by $(N)$: $(\beta_0)$ stands for an $SU(\beta_0)$ gauge node, $(\beta_0 + \beta_1)$ for an $SU(\beta_0 + \beta_1)$ gauge node, and so on. There are two rows of gauge groups for the quiver in Figure \ref{quiver}. Associated with each gauge node there is a $(4, 4)$ vector multiplet. $SU(F)$ flavour groups are denoted by $[F]$. Blue lines represent $(4, 4)$ hypermultiplets. They transform in the bifundamental representation of the groups they are attached to. Red lines represent $(0,4)$ hypermultiplets, while dashed lines are $(0,2)$ Fermi multiplets. They also carry one fundamental and one anti-fundamental index of the groups they are attached to. The $F$'s and $\tilde{F}$'s are not independent of the other numbers of the quiver: as noticed in \cite{Lozano:2019zvg} the theory is chiral and might suffer from gauge anomalies. The $F$'s and $\tilde{F}$'s can be chosen in such a way gauge anomalies cancel out at each gauge node of the quiver. A straightforward calculation (see \cite{Lozano:2019zvg}) leads to
\begin{equation}
F_{k-1} = \nu_{k-1} - \nu_k \, , \quad \tilde{F}_{k-1} = \beta_{k-1} - \beta_k \, .
\end{equation}

\section{Spin 2 fluctuations on $AdS_3 \times S^2 \times \text{CY}_2 \times \mathcal{I}_{\rho}$}\label{section fluctuations}

\paragraph{} In this section we aim at studying massive spin 2 fluctuations of the $AdS_3$ metric in \eqref{eq:class I background}. As we will see, they are composed of a transverse, traceless part along the $AdS_3$ direction and a scalar mode along the internal manifold. The goal of this section is to find the equation that such a metric fluctuation should solve. Explicit solutions will be given in the following sections.
\vspace{3pt}

As mentioned in the introduction, the study of the full KK-spectrum of the warped $AdS_3$ background in \eqref{eq:class I background} and \eqref{eq:class I background RR} is not an easy task. However, in \cite{Bachas:2011xa} it has been shown that, in the case of a warped $AdS_4$, the equations for the fluctuation of the metric decouple from all other fluctuations. Moreover, they solve a ten dimensional Laplace equation which depends only on the background metric (such equation will be given later in this section). The analysis done in \cite{Bachas:2011xa} can be extended to any warped background with an $AdS$ factor, and it is  straightforward to apply it to the case we are interested in, namely spin 2 fluctuations of the warped $AdS_3 \times S^2 \times \text{CY}_2 \times \mathcal{I}_{\rho}$.

\subsection*{Equation for spin 2 fluctuations}

\paragraph{} To begin with, let us consider the background metric of \eqref{eq:class I background} in the Einstein frame. This is achieved, as usual, by multiplying the ``string frame'' metric of \eqref{eq:class I background} by $e^{-\Phi/2}$, being $\Phi$ the dilaton. A useful and compact form for it is
\begin{equation}
ds^2 = f_1 e^{- \Phi/2} ds^2_{AdS_3} + \hat{g}_{a b} dz^a dz^b \, ,
\end{equation}
where the warping factor $f_1$ and the ``internal'' metric are given by the following expressions
\begin{equation}
f_1 = \frac{u}{\sqrt{h_4 h_8}} \, , \quad \hat{g}_{a b} dz^a dz^b = e^{-\Phi/2} \bigg(  \frac{u \sqrt{h_4 h_8}}{4 h_4 h_8 + (u')^2} ds^2(S^2) + \sqrt{\frac{h_4}{h_8}} ds^2(\text{CY}_2) + \frac{\sqrt{h_4 h_8}}{u} d \rho^2 \bigg) \, .
\end{equation}
In the following we will take the CY to be a four-torus T$^4$ parametrised by $4$ angles $\theta_i, (i = 1, \dots , 4)$. Here, of course, $\theta_i \cong \theta_i + 2 \pi$. Let us then consider a symmetric fluctuation $h$ along the $AdS_3$ part of the ten-dimensional metric
\begin{equation}
ds^2 = f_1 e^{- \Phi/2} (ds^2_{AdS_3} + h_{\mu \nu} dx^{\mu} dx^{\nu}) + \hat{g}_{a b} dz^a dz^b \, .
\end{equation}
$h$ can be decomposed into a transverse traceless fluctuation on $AdS_3$ and a mode on the internal manifold in the following manner
\begin{equation}\label{eq:factorisation of h}
h_{\mu \nu}(x,z) = h^{[tt]}_{\mu \nu}(x) \psi(z). 
\end{equation}
Following \cite{Bachas:2011xa}, the transverse traceless fluctuation $h^{[tt]}_{\mu \nu}(x)$ satisfies the following equation of motion on $AdS_3$
\begin{equation}\label{eq:equation fluctuation AdS}
\square_{AdS_3}^{(2)} h^{[tt]}_{\mu \nu}(x) = (M^2 -2) h^{[tt]}_{\mu \nu}(x) \, ,
\end{equation}
where $\square_{AdS_3}^{(2)}$ is the Laplace operator acting on massive rank-two tensors in $AdS_3$, see e.g. \cite{Polishchuk:1999nh}. The authors of \cite{Bachas:2011xa} have shown that the linearised Einstein equations reduce to the ten dimensional Laplace equation
\begin{equation}
\frac{1}{\sqrt{-g}} \partial_M \sqrt{-g} g^{MN} \partial_N h_{\mu \nu} = 0 \, .
\end{equation}
For the background metric in \eqref{eq:class I background}, and with $h_{\mu \nu}^{[tt]}$ satisfying the equation \eqref{eq:equation fluctuation AdS}, we get the following equation for the ``internal mode'' $\psi(z)$
\begin{equation}\label{equation internal manifold}
\frac{(f_1 e^{- \Phi/2})^{-1/2}}{\hat{g}^{1/2}} \partial_a \left[ (f_1e^{- \Phi/2} )^{3/2} \sqrt{\hat{g}} \hat{g}^{ab} \partial_b \right] \psi(z) = -M^2 \psi(z) \, .
\end{equation}
Expanding out equation \eqref{equation internal manifold} we find
\begin{equation}\label{equation internal manifold expanded}
\left[ \bigg( 4 + \frac{(u')^2}{\hat{h}_4 h_8} \bigg) \nabla^2_{S^2} + \frac{u}{\hat{h}_4} (\partial^2_{\theta_1} + \partial^2_{\theta_2^2} + \partial^2_{\theta_3} + \partial^2_{\theta_4}) + \frac{1}{\hat{h}_4 h_8} \frac{d}{d \rho} \big( u^2 \frac{d }{d \rho} \big) + M^2 \right] \psi (z) = 0 \, .
\end{equation}
The function $\psi$ can be conveniently decomposed into spherical harmonics on the $S^2$ and into plane waves on the T$^4$ in the following fashion
\begin{equation}\label{decompose internal mode}
\psi = \sum_{l , m, n} \psi_{l m n} Y_{l, m} e^{i n \cdot \theta} \, .
\end{equation}
Here $n$ is a shorthand notation for $(n_1, n_2, n_3, n_4)$ and $n \cdot \theta = n_1 \theta_1 + n_2 \theta_2 + n_3 \theta_3 + n_4 \theta_4$. The $n_i$'s are of course integers, in order for $\psi$ to be single valued. Substituting \eqref{decompose internal mode} into \eqref{equation internal manifold expanded} we get an equation for $\psi_{lmn}$ which reads
\begin{equation}\label{almost Sturm-Liouville equation}
\frac{1}{\hat{h}_4 h_8} \frac{d }{d \rho} \bigg( u^2 \frac{d \psi_{lmn}}{d \rho} \bigg) - \left[ \bigg( 4 + \frac{(u')^2}{\hat{h}_4 h_8} \bigg) l(l+1) + \frac{u}{\hat{h}_4} n^2 - M^2 \right] \psi_{lmn}  = 0 \, .
\end{equation}

It turns out to be useful to redefine $\psi_{l m n} = u^l \phi_{l m n}$. In this way, equation \eqref{almost Sturm-Liouville equation} becomes an equation for $\phi_{lmn}$, which reads\footnote{To get the equation \eqref{Sturm-Liouville equation}, we need to use $u'' = 0$, which is globally true.}
\begin{equation}\label{Sturm-Liouville equation}
\frac{d}{d \rho} \left( u^{2(l+1)} \frac{d \phi_{lmn}}{d \rho} \right) - n^2 \hat{h}_4 h_8 u^{2 l} \left( \frac{u}{\hat{h}_4} \right) \phi_{lmn} = - (M^2 - 4l (l+1)) \hat{h}_4 h_8 u^{2 l} \phi_{lmn} \, ,
\end{equation}
or, in a more compact form,
\begin{equation}\label{Sturm-Liouville problem}
S \phi_{l m n} + q(\rho) \phi_{l m n} = - \lambda w(\rho) \phi_{l m n} \, ,
\end{equation}
with the differential operator $S$ and the functions $q$ and $w$ given by
\begin{equation}\label{Sturm-Liouville functions}
S = \frac{d}{d \rho} \left( p(\rho) \frac{d}{d \rho} \right) \, ,  \quad p(\rho) = u^{2(l+1)} \, , \quad q(\rho) = - n^2 \hat{h}_4 h_8 u^{2 l} \left( \frac{u}{\hat{h}_4} \right) \, ,\quad w(\rho) = \hat{h}_4 h_8 u^{2 l} \, ,
\end{equation}
while the ``eingenvalue'' $\lambda$ is
\begin{equation}
\lambda = M^2 - 4l (l+1) \, .
\end{equation}
The equation \eqref{Sturm-Liouville problem}, together with the definitions \eqref{Sturm-Liouville functions}, defines a \textit{Sturm-Liouville problem}\footnote{The three functions $u$, $\hat{h}_4$ and $h_8$ are, of course, always positive definite, in order for the background metric in \eqref{eq:class I background} to have the correct signature. Therefore $w(\rho)$ is always positive definite. This condition is necessary to have a well defined Sturm-Liouville problem.}. As we discussed in Section \ref{section 2}, the variable $\rho$ takes values in the finite interval $\mathcal{I}_{\rho} = [0 , 2 \pi (P+1)]$ and the function $u$ vanishes only at $\rho = 0$. Therefore we have what in the mathematical literature is known as a singular Sturm-Liouville problem.

Notice also that with the substitution $d \rho / d t = u^{2(l+1)}$, with $t$ a new variable, the equation \eqref{Sturm-Liouville equation} reduces to a Schr\"{o}dinger-like equation. We will not be studying \eqref{Sturm-Liouville equation} in its Schr\"{o}dinger form. Our starting point will be equation \eqref{Sturm-Liouville equation} and, as we will see in coming sections, solutions to that equation can be found.

\section{Unitarity and a special class of solutions}\label{section universal solution}

\paragraph{} In this section we will show how a bound for $M^2$ emerges from equation \eqref{Sturm-Liouville equation}. For this bound, we find a particular class of solutions which will be dubbed ``universal". Regularity conditions for the mode $\psi$ will also be discussed.

\subsection*{A bound for $M^2$} 

\paragraph{} To begin with, let us multiply \eqref{Sturm-Liouville equation} by $\phi_{lmn}$ and then integrate over $\rho$. The equation we get is 
\begin{equation}
\int_{\mathcal{I}_{\rho}} d \rho \, \phi \frac{d}{d \rho} \left( u^{2(l+1)} \frac{d \phi}{d \rho} \right) - n^2 \hat{h}_4 h_8 u^{2 l} \left( \frac{u}{\hat{h}_4} \right) \phi^2 + (M^2 - 4l (l+1)) \hat{h}_4 h_8 u^{2 l} \phi^2 = 0 \, ,
\end{equation}
where $\phi$ stands for $\phi_{lmn}$. Now, if we integrate by parts the first term we get
\begin{equation}\label{equation defining bound}
\int_{\mathcal{I}_{\rho}} d \rho \left( - \phi'^2  u^{2(l+1)} - n^2 h_8 u^{2 l + 1} \phi^2 + (M^2 - 4l (l+1)) \hat{h}_4 h_8 u^{2 l} \phi^2 \right)= - \phi \phi' u^{2(l+1)}\Big\rvert_{0}^{\rho^{*}} \, .
\end{equation}
Notice that $\phi \phi' u^{2(l+1)}$ vanishes when evaluated at $\rho = 0$ (as $u$ does vanish at $\rho = 0$) as long as $\phi$ and $\phi'$ are regular there, while it doesn't when evaluated at $\rho = \rho^*$. In the following, we will focus on the Hilbert space of functions $\phi$ for which $\phi \phi' u^{2(l+1)}$ vanishes also at $\rho = \rho^*$. Thus, the equation \eqref{equation defining bound} reduces to
\begin{equation}
\int_{\mathcal{I}_{\rho}} d \rho \left( \phi'^2  u^{2(l+1)} + n^2 h_8 u^{2 l + 1} \phi^2 \right) = (M^2 - 4l (l+1)) \int_{\mathcal{I}_{\rho}} d \rho \, \hat{h}_4 h_8 u^{2 l} \phi^2 \, ,
\end{equation}
Given that $u$, $\hat{h}_4$ and $h_8$ are non-negative, and the integrals finite, we find the following lower bound for $M^2$
\begin{equation}\label{eq:bound}
M^2 \geq 4l(l+1) \, .
\end{equation}

\subsection*{Universal minimal solution}

\paragraph{} Let us consider the case where $M^2 = 4l(l+1)$ and $n = 0$. Then, equation \eqref{Sturm-Liouville equation} simply reduces to
\begin{equation}\label{easiest equation}
\frac{d}{d \rho} \Big( u^{2(l+1)} \frac{d \phi_{lm}}{d \rho} \Big) = 0 \, ,
\end{equation}
which can be integrated to give $\phi'_{lm} = \text{constant}/u^{2(l+1)}$. However, for the class of geometries discussed in Section \ref{section 2}, $u$ vanishes at $\rho = 0$ (it is in fact linear in $\rho$) and therefore $\phi'_{lm}$ is not finite at $\rho = 0$. This, in turn, implies that both $\phi_{lm}$ and $\psi_{lm} = u^l \phi_{lm}$ are not finite at $\rho = 0$ for any $l$. As we are looking for fluctuations that remain finite everywhere, the only acceptable solution to \eqref{easiest equation} is $\phi_{lm} = \text{constant}$. This in turn implies that
\begin{equation}\label{eq:minimal solution}
\phi_{l m} = \text{constant} \, , \quad \quad \quad \psi_{lm} = \text{constant} \times u^l \, , \quad \quad \quad M^2 = 4l(l+1) \, .
\end{equation}
This class of solutions is independent of the form of $u$, $\hat{h}_4$ and $h_8$ and in this sense they are ``universal". Moreover, they are the solutions with minimal $M$ for a given $l$, saturating the bound \eqref{eq:bound}, and therefore correspond to ``minimal" solutions.

The bound \eqref{eq:bound} for the mass of spin 2 excitations will prove to be very important when discussing quantum field theory implications. In particular, anticipating the discussion in Section \ref{section implications for field theory}, the spin 2 fluctuations considered are dual to operators in the field theory with dimension $\Delta$ given by the usual AdS/CFT formula, $M^2 = \Delta(\Delta -2)$. The inequality \eqref{eq:bound} implies for the conformal dimension the following lower bound
\begin{equation}\label{eq:bound on the dimension}
\Delta \geq \Delta_{\text{min}} \, ,  \quad \quad \quad \Delta_{\text{min}} = 2l + 2 \, .
\end{equation}
We leave a further discussion of the bound \eqref{eq:bound on the dimension} for later.

For the case of non universal solutions, i.e. solutions for which $M^2 > 4l(l+1)$, it is necessary to specify what the three functions $u$, $\hat{h}_4$ and $h_8$ are. The general form of these functions has been given in Section \ref{section 2}. An example of non universal solution to \eqref{Sturm-Liouville equation} will be discussed in the Appendix \ref{section non universal solution}. 

\section{Implications for the dual field theory}\label{section implications for field theory}

\paragraph{} In this section we identify the operators dual to the spin 2 fluctuations that we have studied in the previous sections. Crucial for this would be the comparison of the spectrum of fluctuations with the spectrum of multiplets built in \cite{Kos:2018glc}. The analysis of \cite{Kos:2018glc} uses insights developed in \cite{Cordova:2016emh, Cordova:2016xhm} for the construction of supermultiplets in $d \geq 3$ and is sketched in Appendix \ref{section superconformal algebra}.

\subsection*{Superconformal multiplets}

\paragraph{} For each fluctuation of the metric introduced above there is an operator in the dual SCFT. Therefore, we should aim at understanding what kind of operators these metric fluctuations correspond to. As already mentioned, the representations of $\mathcal{N} = 4$ superconformal algebra in two dimensions were worked out in \cite{Kos:2018glc} and briefly sketched in Appendix \ref{section superconformal algebra}. These representations are labelled by the conformal weight $h$ and $\tilde{h}$ of the $SL(2, \mathbb{R}) \times SL(2, \mathbb{R})$ conformal algebra and the Dynkin index $r$ of the $SU(2)_R$ R-symmetry. In particular, the scaling dimension $\Delta$ of any operator is usually given as the sum of its conformal weights, $\Delta = h +\tilde{h}$. The spin of such operators is determined as the difference between their conformal weights, $s = h - \tilde{h}$. Thus, a state in the superconformal algebra can be represented schematically as
\begin{equation}
[h, \tilde{h}]_{\Delta = h+ \tilde{h}}^{(r)} \, .
\end{equation}

The $SU(2)_R$ is realised on the supergravity side as the isometries of $S^2$. Thus, the quantum number on the $S^2$, $l$ and $m$, are related to the R-charges of the corresponding dual operators. In particular, the Dynkin $r$, which is always an integer, is related to $l$ by $r = 2l$. Therefore, in our construction $r$ will always be a positive, even-integer.

As noted earlier, the mass of a spin 2 bulk field and the scaling dimension of the dual operator are related by the formula $M^2=\Delta(\Delta - 2)$. Thus, the minimal solution \eqref{eq:minimal solution}, for which $M^2 = 4l(l+1)$, corresponds to operators with scaling dimension $\Delta = 2l + 2$. Finally, we should stress that, for the type of fluctuations that we are studying, $h$ and $\tilde{h}$ are not really independent. The $SL(2, \mathbb{R}) \times SL(2, \mathbb{R})$ isometry of $AdS_3$ (plus gauge invariance) classifies $h_{\mu \nu}$ to have $h - \tilde{h} = \pm 2$.

Having identified all the quantum numbers using the standard holographic map, the (complex) spin 2 fluctuations correspond to operators labelled as
\begin{equation}\label{holographiclabel}
[h, h \pm 2]^{(2l)}_{\Delta = 2h \pm 2 = 2l + 2} \, .
\end{equation}
For $h = 2$, \eqref{holographiclabel} comprises of $[2,0]_{\Delta = h = 2}^{(0)}$. These are the quantum numbers of the holomorphic stress energy tensor. As explained in \cite{Kos:2018glc} and in Appendix \ref{section superconformal algebra}, such a state arises as top component descendant in the short multiplet whose conformal primary has $r = 2l = 2$ and $h = r/2 = 1$. Notice also that choosing $h = 0$, \eqref{holographiclabel} leads to $[0,2]_{\Delta = \tilde{h} = 2}^{(0)}$, the quantum numbers of the anti-holomorphic stress energy tensor.

For massive solutions with $l>0$ the spin 2 universal fluctuations \eqref{eq:minimal solution} either correspond to operators which sit as top components in a multiplet whose primary field has dimension $\Delta = 2l+1$, very much as explained in \cite{Kos:2018glc} and in Appendix \ref{section superconformal algebra}, or to operators obtained by tensoring chiral primaries in short multiplets with the anti-holomorphic sector of the algebra, just like the anti-holomorphic energy momentum tensor above.

It would be nice to understand how this operators are built from the fields of the SCFT at hand (the SCFTs represented by the quiver in Figure \ref{quiver}). More in particular, we expect the operators dual to \eqref{eq:minimal solution} to be given by single traces of products of elementary fields in our SCFT. A step forward for this would be to identify the scalar primary $\mathcal{T}$ in the stress-energy tensor multiplet. This, in turn, can be ``dressed'' by other fields in order to get an operator whose scaling dimension $\Delta$ is equal to $2l+1$ and whose R-charge under the $SU(2)_R$ symmetry is $2l+2$. However, we should also take into account that the SCFTs at hand are inherently strongly coupled and a Lagrangian description for them might not be suitable.

\section{Central charge from the spin 2 fluctuations}\label{section central charge}

\paragraph{} In this section we will briefly show a possible way to compute the central charge for the theories in \eqref{eq:class I background}, \eqref{eq:class I background RR}. To this end, we should compute the normalisation of the two-point function of the operators dual to the graviton fluctuations studied in Section \ref{section universal solution}. We have seen in Section \ref{section universal solution} that the universal, minimal solution with $l=m=n=0$ corresponds to a massless graviton and, therefore, the dual operator is the energy momentum tensor. The normalisation of the two-point function for the energy momentum tensor is read off from the effective action for the three-dimensional graviton.

Let us start from the type IIA action written schematically in the Einstein frame as
\begin{equation}
S_{\text{IIA}} = \frac{1}{2 \kappa_{10}^2} \int d^{10} x \sqrt{-g} R + \cdots \, .
\end{equation}
Expanded to second order, and following \cite{Gutperle:2018wuk}, it leads to an action for $h_{\mu \nu}$ which reads
\begin{equation}\label{eq:action for h}
S[h] = \frac{1}{\kappa_{10}^2} \int d^{10}x \sqrt{-g} \, {h}^{\mu \nu} \frac{1}{\sqrt{-g}} \partial_M \sqrt{-g} g^{MN} \partial_{N} h_{\mu \nu} + \text{boundary term}\, .
\end{equation}
Expanding out \eqref{eq:action for h} and dropping the boundary term which is not necessary in what follows, we get
\begin{equation}\label{eq:action for h expanded}
S[h] = \frac{1}{\kappa_{10}^2} \int d^{10}x (-g_{AdS_3})^{\frac{1}{2}} (\hat{g})^{\frac{1}{2}} (f_1 e^{-\frac{\Phi}{2}})^{\frac{1}{2}} \, {h}^{\mu \nu} \left\{ \square_{AdS_3}^{(2)} + 2 +\hat{\square}\right\} h_{\mu \nu} \, ,
\end{equation}
where $\hat{\square}$ is the operator on the left-hand side of \eqref{equation internal manifold}. Using the Ansatz\footnote{Notice that we are using the subscripts $l$, $m$ and $n$ under $h$. This is because in some solutions (like the ``universal'' solution above) $M^2$ depends on those quantum numbers and so does $h^{[tt]}$ through equation \eqref{eq:equation fluctuation AdS}.} $h_{\mu \nu} = (h_{lmn}^{[tt]})_{\mu \nu} Y_{lm} \psi_{lmn} e^{i n \cdot \theta} $ we find
\begin{equation}
S[h] = \sum_{lmn} C_{lmn} \int d^{3}x  \sqrt{-g_{AdS_3}} \, {(h^{[tt]}_{lmn})}^{\mu \nu} \left\{ \square_{AdS_3}^{(2)} + 2 -M^2 \right\} {(h^{[tt]}_{lmn})}_{\mu \nu}
\end{equation}
where the coefficients $C_{lmn}$ are given by\footnote{Using the standard normalisation $\int Y_{lm} Y_{l' m'} = \delta_{l l'} \delta_{m m'}$.}
\begin{equation}\label{eq:normalisation of the spin 2}
C_{lmn} = \frac{16 \pi^4}{\kappa_{10}} \int_{\mathcal{I}_{\rho}} d \rho \sqrt{\hat{g}} \, (f_1 e^{- \frac{\Phi}{2}})^{\frac{1}{2}} |\psi_{lmn}|^2 \, .
\end{equation}

The integral in \eqref{eq:normalisation of the spin 2} is finite for the class of solutions discussed in this paper, namely those fluctuations that are finite everywhere. In particular, if we specialise to the universal, minimal solution \eqref{eq:minimal solution} with $l=m=n=0$, i.e. $\psi_{lmn} = 1$, \eqref{eq:normalisation of the spin 2} evaluated on \eqref{eq:class I background} gives\footnote{Even though, in general AdS backgrounds, gravitons should be treated separately from massive spin 2 fields (see for instance \cite{Arutyunov:1998hf}), the normalisation for the three dimensional massless graviton can still be obtained by taking $l = m = n = 0$}
\begin{equation}\label{eq:central charge}
C_{0} = \frac{1}{4 \kappa_{10}^2} \text{Vol(CY$_2$)} \int_{\mathcal{I}_{\rho}} d \rho \, \hat{h}_4 h_8 \, .
\end{equation}
The effective three-dimensional gravitational coupling $\kappa_3$ is related to $C_0$ by  $C_0 = 1/\kappa_3^2$. The quadratic action for $h_{\mu \nu}$ computes the two point function of the dual stress tensor, whose coefficient is well known to be proportional to the central charge of the 2d CFT. In fact, \eqref{eq:central charge} is equal, modulo an irrelevant numerical factor, to the central charge computed on pag. 12 of \cite{Lozano:2019zvg}.

\section{Conclusions}\label{section conclusions}

\paragraph{} In this paper we have investigated aspects of spin 2 fluctuations around the background $AdS_3 \times S^2 \times \text{CY}_2 \times \mathcal{I}_{\rho}$ of \cite{Lozano:2019emq}.
An equation for these spin 2 fluctuations has been derived in Section \ref{section universal solution}, following the general analysis of \cite{Bachas:2011xa}, and we have seen that they fall into two classes, universal and non universal solutions.

The universal solutions, discussed in Section \ref{section universal solution}, turned out to be particularly interesting, as they are independent of the background data. These fluctuations, and therefore the dual operators, are expected to be present for any of the backgrounds given in \eqref{eq:class I background} and \eqref{eq:class I background RR}.  As we have seen in the main text, they are dual to operators with scaling dimension $\Delta = 2l+2$, where $l$ is the angular-momentum-charge on the $S^2$ which realises holographically the $SU(2)_R$ symmetry of the dual field theory. 

The non universal solutions are more difficult to analyse as they depend on background data, namely on a specific choice of the functions $u$, $\hat{h}_4$ and $h_8$ given in \eqref{eq:h4}, \eqref{eq:h8}, \eqref{eq:u}. An example of these is worked out in Appendix \ref{section non universal solution}. 

Finally, we have seen in Section \ref{section central charge}  that the central charge $c$ for the 2d dual quiver field theory can be read off from the normalisation of the action for the spin 2 fluctuations, $h_{\mu \nu}$. The central charge $c$ is essentially determined from the 3-dimensional gravitational coupling constant $\kappa_3$, $c \propto \kappa_3^{-2}$. The quadratic actions for $l > 0$ modes compute the two point functions for the corresponding dual operators and, more in particular, the holographic normalisations of these operators. Moreover, computing higher order interaction terms, one can then compute three point and higher point functions of these operators. 

\section*{Acknowledgements}

\paragraph{} A special thank goes to Carlos Nunez for constant help and discussion throughout the completion of this work. I am also very grateful to Adi Armoni, Prem Kumar, Yolanda Lozano, Daniel Thompson and Salomon Zacarias for stimulating discussions and correspondence. I would like to thank Mohammad Akhond and Giacomo Piccinini for a careful reading of the manuscript and comments.
\appendix

\section{Example of non universal solution}\label{section non universal solution}

\paragraph{} In this appendix we consider a particular solution to \eqref{Sturm-Liouville equation} which is not universal, namely a solution that does not saturate the bound $M^2 = 4l(l+1)$, still with $n=0$. In order to solve \eqref{Sturm-Liouville equation}, we will have to choose some particular $u$, $\hat{h}_4$ and $h_8$ which, in turn, correspond to a particular background. Let us start off by considering the case of\footnote{This is of course a particular example of equations \eqref{eq:h4}, \eqref{eq:h8}, \eqref{eq:u}.}
\[ \hat{h}_4(\rho) = \beta_0 \begin{cases} 
      \rho/2 \pi & 0 \leq \rho \leq  \pi P\\
      P - \rho/2 \pi &   \pi P < \rho \leq 2 \pi P
   \end{cases} \, , \quad \quad h_8 = \hat{h}_4 \, , \quad \quad u = \frac{\beta_0}{2 \pi} \rho \, .
\]
A solution to \eqref{Sturm-Liouville equation} must be split into two solutions in the two intervals $ \mathcal{I}_{I} = \left[0,  \pi P \right]$ and $\mathcal{I}_{II}=\left(\pi P, 2 \pi P \right]$, as both $\hat{h}_4$ and $h_8$ are only piecewise continuous. Moreover, in order to get a smooth solution for the fluctuations, we need to impose continuity of the solution and of its derivative at $\rho = \pi P$.

Equation \eqref{Sturm-Liouville equation} for $u$, $\hat{h}_4$, $h_8$ as before  looks like
\begin{equation}\label{eq:equation region I}
\begin{split}
&\phi''(\rho) + \frac{2 l + 2}{\rho} \phi'(\rho) + \lambda \phi(\rho) = 0  \quad \quad \quad \quad \quad \quad \quad \quad \text{in} \quad \mathcal{I}_I \, , \\
&\phi''(\rho) + \frac{2 l+2}{\rho} \phi'(\rho) + \lambda \frac{ (P-\rho/2 \pi)^2}{\rho^2} \phi(\rho) = 0 \quad \quad \, \, \, \text{in} \quad \mathcal{I}_{II} \, ,
\end{split}
\end{equation}
where again $\lambda = M^2 - 4l(l+1)$. The general solution of \eqref{eq:equation region I} in $\mathcal{I}_I$ reads $\phi = c_1 \phi_1(\rho) + c_2 \phi_2(\rho)$, with 
\begin{equation}
\phi_1 = \rho^{ -\frac{2l+1}{2}} \, J_{\frac{2l+1}{2}} \big( \sqrt{\lambda} \rho \big) \quad \text{and} \quad \phi_2 = \rho^{ -\frac{2l+1}{2}} \, Y_{\frac{2l+1}{2}} \big( \sqrt{\lambda} \rho \big) \, .
\end{equation}
$J$ and $Y$ are Bessel functions of the first and second kind, respectively and $c_1$ and $c_2$ are integration constants. In order for the solution (and its derivative) to be regular at $\rho = 0$ we must set $c_2 = 0$. On the other hand, the general solution to \eqref{eq:equation region I} in $\mathcal{I}_{II}$ can be given in terms of the complex function $\phi = \tilde{c}_3 \tilde{\phi}_3 + \tilde{c}_4 \tilde{\phi}_4$, with
\begin{equation}
\begin{split}
\tilde{\phi}_3 &= e^{-i \sqrt{\lambda} \frac{\rho}{2 \pi}} \rho^{- (l + 1/2 - i \frac{\gamma}{2})} U \big( \alpha, \gamma , i \sqrt{\lambda} \rho/\pi \big)  \\
\tilde{\phi}_4 &= e^{-i \sqrt{\lambda} \frac{\rho}{2 \pi}} \rho^{- (l + 1/2 - i \frac{\gamma}{2})} M \big( \alpha, \gamma , i \sqrt{\lambda} \rho/\pi \big) \, ,
\end{split}
\end{equation}
where $U$ and $M$ are the Kummer's hypergeometric functions, respectively, and $\alpha$ and $\gamma$ are two complex numbers given by\footnote{The argument of the square root appearing in the definition of $\gamma$ is positive in the limit of very large $P$. This is the regime well described by supergravity. Moreover, for generic $M$ and  $l$ with $M^2 > 4l(l+1)$, $\gamma$ is never a negative integer. Thus the Kummer's functions are always well defined.}
\begin{equation}
\alpha = \frac{\gamma}{2} - i \sqrt{\lambda} \, , \quad \quad \quad \gamma = 1 + i \{4 M^2 P^2 - 4l(l+1) (1 + 4 P^2) - 1 \}^{1/2} \, .
\end{equation}

The functions $\tilde{\phi}_3$ and $\tilde{\phi}_4$ are always well defined in $\mathcal{I}_{II}$. Therefore neither $\tilde{c}_3$ nor $\tilde{c}_4$ must be set to zero. Moreover, we can always consider two independent real combinations of $\tilde{\phi}_3$ and $\tilde{\phi}_4$. Let us call them $\phi_3$ and $\phi_4$. Thus, in $\mathcal{I}_{II}$ the general solution reads $\phi = c_3 \phi_3 + c_4 \phi_4$, with $\phi_3$ and $\phi_4$ two real linearly independent functions built from $\tilde{\phi}_3$ and $\tilde{\phi}_4$ above.

We should now match the solution $\phi = c_1 \phi_1$ with $\phi = c_3 \phi_3 + c_4 \phi_4$ at $\rho = \pi P$. This leads to two conditions
\begin{equation}\label{homogeneous 1}
\phi \big\vert_{\pi P^-} = \phi \big\vert_{\pi P^+} \, , \quad \phi' \big\vert_{\pi P^-} = \phi' \big\vert_{\pi P^+}
\end{equation}
As a further condition, we would like to impose that either $\phi$ or $\phi'$ vanishes at $\rho = 2 \pi P$. This is nothing but the condition discussed around \eqref{equation defining bound}. Say, for instance, that is $\phi$ that vanishes at $\rho = 2 \pi P$
\begin{equation}\label{homogeneous 2}
\phi \big\vert_{2 \pi P} = 0 \, .
\end{equation}
We therefore get a system of three equations, \eqref{homogeneous 1} and \eqref{homogeneous 2}, for three integration constants, $c_1$, $c_3$ and $c_4$. A straightforward calculation shows that such a system has a non trivial solution if and only if the following equation is satisfied
\begin{equation}
\det M = 0 \, \quad \text{with} \quad M = \begin{pmatrix} 
\phi_1 \big\vert_{\pi P} & - \phi_3 \big\vert_{\pi P} & -  \phi_4 \big\vert_{\pi P} \\
\phi'_1 \big\vert_{\pi P} & - \phi'_3 \big\vert_{\pi P} & - \phi'_4 \big\vert_{\pi P} \\
0 & \phi_3 \big\vert_{2 \pi P} &  \phi_4 \big\vert_{2 \pi P} 
\end{pmatrix}\, .
\end{equation}
Such an equation could be solved numerically for $M^2$. Even though we will not attempt at solving it, the expectation is to find a solution of the form
\begin{equation}
M^2 = 4l (l+1) + j f(j, l) \, , \quad  j \in \mathbb{Z}_{\geq 0} \, ,
\end{equation}
where $f$ is a generic positive function of $j$ and $l$ such that $f(0, l)$ is regular.

\section{$\mathcal{N} = 4$ superconformal algebra}\label{section superconformal algebra}

\paragraph{} In this appendix we review the small $\mathcal{N} = 4$ superconformal algebra that was first derived in \cite{Ademollo:1975an}. We will follow Section 2 of \cite{Kos:2018glc}.

The algebra we are considering is a graded Lie algebra with an internal $SU(2)_R$ symmetry and reads
\begin{equation}
\begin{split}
&[L_n, L_m] = (n-m)L_{n+m} + \frac{1}{2} k n (n^2-1) \delta_{n,-m} \, , \\
&\{G^a_r, G^b_s \} = \{\bar{G}^a_r, \bar{G}^b_s \} = 0 \, , \\
&\{G^a_r, \bar{G}^b_s \} = 2 \delta^{ab} L_{r + s} - 2 (r - s) \sigma_i^{ab} J^i_{r + s} + \frac{1}{2} k (4 r^2 -1) \delta_{r, -s} \delta^{ab} \, , \\
&[L_n, J_m^i] = -m J_{n + m} \, , \\
&[L_n, G_r^a] = -\Big( \frac{1}{2}n - r \Big) G_{n + r}^a \, , \\
&[J_n^i, G_r^a] = \frac{1}{2} \sigma^i_{a b} G_{n + r}^b \, , \\
&[J_n^i, J_n^j] = i \epsilon^{ijk} J^k_{n+m} + \frac{1}{2} k n \delta_{n, -m} \, .\\
\end{split}
\end{equation}
Here $L_n$ and $G^a_r$ are the generators of superconformal symmetry. $G^a_r$'s carry an $SU(2)_R$ fundamental index, $a$, and therefore they form an $SU(2)$ complex doublet. $J_n^i$ ($i = 1, 2, 3$) are the $SU(2)_R$ Kac-Moody currents generating the corresponding Kac-Moody loop algebra.  $\sigma_i^{ab}$ are Pauli matrices. Indices $n$ and $m$ run over integer numbers while $r$ belongs to $\mathbb{Z} +1/2$: only for the NS-sector there exists a finite dimensional subalgebra generated by $L_0$, $L_{\pm 1}$, $G_{\pm 1/2}^a$ and $J_0^i$ (see below).

In the following we will mainly be interested in the global part of the superconformal algebra. This reads
\begin{equation}
\begin{split}
&[L_{+1}, L_{-1}] = 2 L_0 \, , \quad [L_{\pm1}, L_{0}] =\pm L_{\pm} \, ,\\
&\{G^a_{\pm \frac{1}{2}}, G^b_{\pm \frac{1}{2}} \} = \{\bar{G}^a_{\pm \frac{1}{2}}, \bar{G}^b_{\pm \frac{1}{2}} \} = 0 \, , \\
&\{G^a_{+ \frac{1}{2}}, \bar{G}^b_{- \frac{1}{2}} \} = 2 \delta^{ab} L_{0} - 2 \sigma_i^{ab} J^i_{0}\, , \quad \{G^a_{\pm \frac{1}{2}}, \bar{G}^b_{\pm \frac{1}{2}} \} = 2 \delta^{ab} L_{\pm 1} \\
&\{G^a_{- \frac{1}{2}}, \bar{G}^b_{+ \frac{1}{2}} \} = 2 \delta^{ab} L_{0} + 2 \sigma_i^{ab} J^i_{0}\, , \\
&[L_0, G_{\pm \frac{1}{2}}^{a}] = \mp \frac{1}{2} G_{\pm \frac{1}{2}}^a \, , \\
&[L_{\pm 1}, G_{\mp \frac{1}{2}}^{a}] = \pm G_{\pm \frac{1}{2}}^a \, , \\
&[J_0^i, G_{\pm \frac{1}{2}}^a] = - \frac{1}{2} \sigma^i_{a b} G_{\pm \frac{1}{2}}^b \, , \\
&[J_0^i, J_0^j] = i \epsilon^{ijk} J^k_{0} \, .\\
\end{split}
\end{equation}

A highest weight state of the superconformal algebra can be specified by the eigenvalues of the mutually commuting operators $L_0$, $\vec{J}_0^2$ and $J_0^3$, $| \mathcal{O}_{h, l} \rangle$, satisfying
\begin{equation}
L_0 | \mathcal{O}_{h, j} \rangle = h | \mathcal{O}_{h, j} \rangle \, , \quad \vec{J}_0^2 | \mathcal{O}_{h, j} \rangle = j(j+1) | \mathcal{O}_{h, j} \rangle  \, \, \quad J_0^3 | \mathcal{O}_{h, j} \rangle = j | \mathcal{O}_{h, j} \rangle \, ,
\end{equation}
as well as
\begin{equation}
L_{n} | \mathcal{O}_{h, j} \rangle = G_{r}^a | \mathcal{O}_{h, j} \rangle = \bar{G}_{r}^a | \mathcal{O}_{h, j} \rangle =  J_{n}^i | \mathcal{O}_{h, j} \rangle =0 \, , \quad \quad n, r > 0 \, .
\end{equation}
The correspondence between the heighest weight state $| \mathcal{O}_{h, j} \rangle$ and the corresponding operator of conformal weight $h$ and $SU(2)_R$ spin $j$ is made as usual $| \mathcal{O}_{h, j} \rangle = \mathcal{O}_{h, j} | 0 \rangle$, where $| 0 \rangle$ is the conformal vacuum. In the following, it will make no difference the use of $\mathcal{O}_{h, j}$ or $ |\mathcal{O}_{h, j} \rangle$.

The operators $G^a_{\frac{1}{2}}$ and $\bar{G}^a_{-\frac{1}{2}}$ can be used to derive the full module $\mathcal{L}_r$ from a superconformal primary state $ |\mathcal{O}_{h, j} \rangle$. Being fermionic operators, they can act on a state $ |\mathcal{O}_{h, j} \rangle$ until they annihilate. Thus the length of a module is finite and determined by Fermi statistics. In deriving the full modules, we should also make sure that the various representations are unitary. This will lead to shortening conditions and constraints on the allowed values of $h$ and $j$, as we now shall see.

\subsection*{Singular vectors, short and long multiplets}

\paragraph{} The superconformal algebra constrains the values the conformal weight $h$ can assume. In particular, in (super)conformal theories unitarity implies a lower bound for the scaling dimension of operators as a function of the other quantum numbers in the algebra. The details of the bound depend of course on the particular theory and corresponding algebra. Let us see how this works in our case\footnote{Unitarity for $\mathcal{N}=2$ and $\mathcal{N}=4$ algebras in two dimensions were also discussed in \cite{Eguchi:1987sm, Eguchi:1987wf, Eguchi:1988af}.}.

Consider a superconformal primary state of conformal weight $h$ and $SU(2)_R$ spin $j$ and the fact that\footnote{Notice that the same conclusion can be reached by sandwiching $\{G^a_{-\frac{1}{2}}, \bar{G}^a_{\frac{1}{2}} \}$.}
\begin{equation}
0 \leq |\bar{G}^a_{-\frac{1}{2}} | \mathcal{O}_{h, j} \rangle |^2 + |G^a_{\frac{1}{2}} | \mathcal{O}_{h, j} \rangle |^2  = \langle \mathcal{O}_{h, j} | \{G^a_{\frac{1}{2}}, \bar{G}^a_{-\frac{1}{2}} \} | \mathcal{O}_{h, j} \rangle \, , \quad \quad \text{no sum over } a \, .
\end{equation}
The superconformal algebra implies
\begin{equation}
0 \leq \langle \mathcal{O}_{h, j} | \{G^a_{\frac{1}{2}}, \bar{G}^a_{-\frac{1}{2}} \} | \mathcal{O}_{h, j} \rangle = \langle \mathcal{O}_{h, j} | 2 L_0 - 2 \sigma_i^{aa} J_0^i| \mathcal{O}_{h, j} \rangle = 2 (h \mp j) \langle \mathcal{O}_{h, j} | \mathcal{O}_{h, j} \rangle \, .
\end{equation}
To have a unitary theory we should then impose $h \geq j$. Sometimes it is customary to use the Dynkin index ($r$ say) of the representation of the internal group $SU(2)_R$. In our case it is related to the spin $j$ by $r= 2 j$. In particular $r$ is always an integer. This is the convention that has been used in \cite{Kos:2018glc}.

Thus, the algebra implies a lower bound for the conformal weight $h$ in terms of the other quantum number $j$. When $h = j$, the module gets shortened as there are null-states that need to be modded out. In particular, when the bound is satisfied there are two states that satisfy
\begin{equation}
\bar{G}_{-\frac{1}{2}}^1| \mathcal{O}_{h, j} \rangle = G_{-\frac{1}{2}}^2 | \mathcal{O}_{h, j} \rangle = 0 \, .
\end{equation}
Therefore, only $\bar{G}_{-\frac{1}{2}}^2$ and $G_{-\frac{1}{2}}^1$ will produce new states. Multiplets of this kind are \textit{short}.

Following \cite{Kos:2018glc}, we just state the result of considering $h = j$ when $j = 1$ (or equivalently $r = 2$). This is the case of most relevance for our purposes as it will lead us to identify the supermultiplet to which the holomorphic energy momentum tensor belongs. A state with $h = j=1$ can be labelled as $[h]^{(j)} = [1]^{(1)}$. The structure of the resulting short multiplet is
\begin{equation*}
\begin{tikzpicture}
\node at (0,0) (1) {$[1]^{(1)}$};
\node at (2,1) (2) {$\left[\frac{3}{2}\right]^{\left(\frac{1}{2}\right)}$};
\node at (2,-1) (3) {$\left[\frac{3}{2}\right]^{\left(\frac{1}{2}\right)}$};
\node at (4,0) (4) {$[2]^{(0)}$};

\draw[->,thick] (1)--(2) ; 
\draw[->,thick] (1)--(3) ;
\draw[->,thick] (2)--(4) ;
\draw[->,thick] (3)--(4) ;   
\end{tikzpicture}
\end{equation*}
where $\nearrow$ stands for the action of $G_{-\frac{1}{2}}$, while $\searrow$ stands for the action of $\bar{G}_{-\frac{1}{2}}$.

As noticed in \cite{Kos:2018glc}, the top component $[2]^{(0)}$ corresponds to the holomorphic energy momentum tensor.

Let us conclude by briefly mentioning the case where $h > j$. In this case there are no null-states and the supermultiplets do not get shortened. All the $G$'s and $\bar{G}$'s contribute to produce new states from the corresponding conformal primary. Such multiplets are therefore \textit{long}. We will not discuss long multiplets any further. A careful analysis can be found in \cite{Kos:2018glc}.

\section{$\mathcal{N} = (0,2)$ and $\mathcal{N} = (0,4)$ theories}\label{2d theories}

\paragraph{} In this appendix we review some basic facts about $\mathcal{N} = (0,4)$ gauge theories. $\mathcal{N} = (0,4)$ superfields are made from $\mathcal{N} = (0,2)$ superfields, therefore we start by reviewing $\mathcal{N} = (0,2)$ gauge theories. For a complete discussion see \cite{Witten:1993yc}.

$\mathcal{N} = (0,2)$ \textbf{multiplets} Let us list the field components of three types of $\mathcal{N} = (0,2)$ multiplets, namely the vector $U$, chiral $\Phi$ and the Fermi $\Psi$ multiplets
\begin{equation}
\boxed{U : (u_{\mu}, \zeta_{-}, D) \, , \quad \quad  \Phi : (\phi, \psi_+) \, , \quad \quad \Psi : (\psi_-, G) \, .}
\end{equation}
The subscript on the fermions refers to their chiralities under $SO(1,1)$ Lorentz group. $D$ is a real and $G$ a complex auxiliary field. The vector $U$ has the following expansion
\begin{equation}
U = u_0 - u_1 - 2 i \theta^+ {\bar{\zeta}}_{-} - 2 i {\bar{\theta}}^+ \zeta_{-} + 2 \theta^+ {\bar{\theta}}^+ D \, .
\end{equation}
The corresponding field strength is formed in the following way
\begin{equation}
\Upsilon = [{\overline{\mathcal{D}}}_+, \mathcal{D}_-] = - \zeta_- - i \theta^+ (D - i u_{01}) - i \theta^+ {\bar{\theta}^+} (\mathcal{D}_0+\mathcal{D}_1) \zeta_- \, ,
\end{equation}
where ${\overline{\mathcal{D}}}_+$ and $\mathcal{D}_-$ are the supercovariant gauge derivatives \cite{Witten:1993yc}. The chiral field $\Phi$ satisfies the following equation
\begin{equation}
{\overline{\mathcal{D}}}_+ \Phi = 0 \, ,
\end{equation}
and therefore expands out in components as
\begin{equation}
\Phi = \phi + \sqrt{2} \theta^+ \psi_- - i \theta^+ {\bar{\theta}}^+ (D_0 + D_1) \phi \, ,
\end{equation}
where $D_0$ and $D_1$ stand for the time- and space-components of the usual covariant derivative.
A Fermi superfield instead satisfies the following equation
\begin{equation}\label{eq:defining fermi multiplet}
{\overline{\mathcal{D}}}_+ \Psi = E (\Phi_i) \, ,
\end{equation}
where $E(\Phi_i)$ is a holomorphic function of the chiral superfields $\Phi_i$. $E$ should be chosen in such a way it transforms as $\Psi$ under all symmetries. Solving \eqref{eq:defining fermi multiplet} leads to the following expansion for $\Psi$
\begin{equation}
\Psi = \psi_- - \sqrt{2} \theta^+ G - i \theta^+ {\bar{\theta}}^+ (D_0 + D_1) \psi_- - \sqrt{2} {\bar{\theta}}^+ E(\phi_i) - 2 \theta^+ {\bar{\theta}}^+ \frac{\partial E}{\partial \phi^i} \psi_{+ i} \, ,
\end{equation}
where $G$ is an auxiliary complex field. The holomorphic function $E$ can be shown to appear in the Lagrangian as a potential term.

There is also another type of superpontential we can consider for $\mathcal{N} = (0,2)$ theories. For each Fermi multiplet $\Psi_a$ we can introduce a holomorphic function $J^a(\Phi_i)$ such that
\begin{equation}
S_J = \int d^2 x d \theta^+ \sum_a J^a(\Phi_i) \Psi_a + \text{h.c.}\, .
\end{equation}
We see that, in analogy to $\mathcal{N} = 1$, $d=4$, $\mathcal{W} = J \cdot \Psi$ is integrated over half superspace.

It must be stressed that the superpotentials $E$ and $J$ cannot be introduced independently without impunity. It turns out that, in order for supersymmetry to be mantained, they have to satisfy
\begin{equation}
E \cdot J = \sum_a E_a J^a = 0 \, .
\end{equation}

Let us now move on to listing $\mathcal{N} = (0,4)$ supermultiplets. They are built from $\mathcal{N} = (0,2)$ supermultiplets.

$\mathcal{N} = (0,4)$ \textbf{multiplets} $\,$ $\mathcal{N} = (0,4)$ supermultiplets are usually given in terms of $\mathcal{N} = (0,2)$ supermultiplets, pretty much as in 4 dimensions $\mathcal{N} = 2$ superfields are built from $\mathcal{N} = 1$ superfields. Again, let us list them first.

\begin{center}
\begin{tabular}{ |c|c|c|c| } 
 \hline
 Multiplets & $\mathcal{N} = (0,2)$ building blocks & component fields& $SU(2)_L \times SU(2)_R$\\ 
  \hline
 Vector & Vector $+$ Fermi $(U, \Theta)$ & $(u_{\mu}, \zeta_{-}^a)$& $(1, 1),(2, 2)$\\ 
 \hline
 Hyper & Chiral $+$ Chiral $(\Phi, \tilde{\Phi})$ & $(\phi^a, \psi_{+}^b)$& $(2, 1),(1, 2)$\\ 
 \hline
 Twisted hyper & Chiral $+$ Chiral $(\Phi', \tilde{\Phi}')$ & $({\phi'}^a, {\psi'}_{+}^b)$& $(1, 2),(2, 1)$\\
  \hline
Fermi & Fermi $+$ Fermi $(\Gamma, \tilde{\Gamma})$ & ${\psi'}_{-}^a$& $(1, 1)$\\
  \hline
\end{tabular}
\end{center}

The $\mathcal{N} = (0,4)$ vector multiplet is made of an $\mathcal{N} = (0,2)$ vector multiplet and an adjoint $\mathcal{N} = (0,2)$ Fermi multiplet $\Theta$. The field content is that of a gauge field $u_{\mu}$ and two left-handed fermions $\zeta_{-}^a$, $a = 1, 2$. The gauge field is a singlet under the $SU(2)_L \times SU(2)_R$ R-symmetry while the two fermions transform as $(2, 2)$.

There are two different types of hypermultiplets, the hypermultiplet and the twisted hypermultiplet. Both of them are formed by two $\mathcal{N} = (0,2)$ chiral multiplets, therefore they both contain two complex scalars ($\phi^a$) and two right-handed fermions ($\psi_{+}^b$). They differ from each other because of the different representations under the R-symmetry group, as we can see from the table above.

If we want to couple the hypermultiplet to the vector multiplet, we should consider the following coupling between the hyper $(\Phi , \tilde{\Phi})$ and the adjoint Fermi field $\Theta$
\begin{equation}
J^{\Theta} = \Phi \tilde{\Phi} \Rightarrow \mathcal{W} = \tilde{\Phi} \Theta \Phi \, .
\end{equation}
This looks very much like the coupling between the hypermultiplet and the chiral adjoint for four dimensional $\mathcal{N} = 2$ theories. On the other hand, coupling a twisted hypermultiplet to the gauge sector requires an E-type of superpotential
\begin{equation}
E_{\Theta} = \Phi' \tilde{\Phi}' \, , 
\end{equation}
with indices in $\Phi' \tilde{\Phi}'$ set to have $E_{\Theta}$ transforming in the adjoint of the gauge group.

Finally, we can have an $\mathcal{N} = (0,4)$ Fermi multiplet, which is made of two $\mathcal{N} = (0,2)$ Fermi multiplets. It contains two left-handed fermions which are singlets of $SU(2)_L \times SU(2)_R$ R-symmetry.
\vspace{3pt}

As the nodes of the quiver of interest, Figure \ref{quiver}, contain $\mathcal{N} = (4, 4)$ vector multiplets and are connected by $\mathcal{N} = (4, 4)$ hypermultiplets, it is probably worth it to mention how $\mathcal{N} = (4, 4)$ superfields decompose.

$\mathcal{N} = (4,4)$ \textbf{multiplets} There are two types of $\mathcal{N} = (4,4)$ superfields, the vector and the hypermultiplet.

\begin{center}
\begin{tabular}{ |c|c|c| } 
 \hline
 Multiplets & $\mathcal{N} = (0,4)$ building blocks & $\mathcal{N} = (0,2)$ building blocks \\ 
  \hline
 Vector & Vector $+$ Twisted Hyper & $(U, \Theta), (\Sigma, \tilde{\Sigma})$ \\ 
 \hline
 Hyper & Hyper $+$ Fermi & $(\Phi, \tilde{\Phi}), (\Gamma, \tilde{\Gamma})$ \\ 
 \hline
\end{tabular}
\end{center}
The $\mathcal{N}=(4,4)$ vector multipled is comprised of an $\mathcal{N}=(0,4)$ vector multiplet and a $\mathcal{N}=(0,4)$ twisted hypermultiplet. The twisted hypermultiplet is usually denoted as $(\Sigma, \tilde{\Sigma})$. They are coupled to the gauge sector via the E-type potential
\begin{equation}
E_{\Theta} = [\Sigma, \tilde{\Sigma}] \, .
\end{equation}
$\mathcal{N} = (4, 4)$ hypermultiplets are made of an $\mathcal{N} = (0, 4)$ hypermultiplet and an $\mathcal{N} = (4, 4)$ Fermi multiplet, all in all $(\Phi, \tilde{\Phi}), (\Gamma, \tilde{\Gamma})$. As before, $\Phi$ and $\tilde{\Phi}$ are coupled to the gauge sector via
\begin{equation}
\mathcal{W} = \tilde{\Phi} \Theta \Phi \, .
\end{equation}

Let us conclude by saying that there are couplings between $\mathcal{N} = (0,4)$ Fermi multiplets $\Gamma$, $\tilde{\Gamma}$, hypermultiplets $\Phi$, $\tilde{\Phi}$ and twisted hypers $\Sigma$, $\tilde{\Sigma}$. They involve both superpotential and E-terms
\begin{equation}
\mathcal{W} = \tilde{\Gamma} \tilde{\Sigma} \Phi + \tilde{\Phi} \tilde{\Sigma} \Gamma \, ,
\end{equation}
and
\begin{equation}
E_{\Gamma} = \Sigma \Phi \, ,\quad E_{\tilde{\Gamma}} = - \tilde{\Phi} \Sigma \, .
\end{equation}
It is easy to see that
\begin{equation}
E \cdot J = \tilde{\Phi} [\Sigma, \tilde{\Sigma}] \Phi + \tilde{\Phi} \tilde{\Sigma} \Sigma \Phi - \tilde{\Phi} \Sigma \tilde{\Sigma} \Phi = 0 \, . 
\end{equation}

\bibliography{Bibliography}
\end{document}